\documentclass[aps,nofootinbib,showpacs,twocolumn,floatfix,superscriptaddress]{revtex4}
\usepackage{graphicx}
\usepackage{epsf,amsfonts,amssymb,amsbsy}
\usepackage[mathscr]{eucal}
\begin{document}
\title{Natural scale of cosmological constant in seesaw mechanism with broken SUSY}
\author{V.V.Kiselev}
 \affiliation{Russian State Research
Center ``Institute for High
Energy Physics'', 
Pobeda 1, Protvino, Moscow Region, 142281, Russia\\ Fax:
+7-4967-744937}
 \affiliation{Moscow Institute of Physics and
Technology, Institutskii per. 9, Dolgoprudnyi, Moscow Region,
141700, Russia}
\author{S.A.Timofeev}
\affiliation{Moscow Institute of Physics and Technology,
Institutskii per. 9, Dolgoprudnyi, Moscow Region, 141700, Russia}
\begin{abstract}
The cosmological constant is inherently determined by the scale of
breaking down supersymmetry in the mechanism of seesaw
fluctuations of two vacuum-states.
\end{abstract}
\maketitle


The evolution of flat Universe is well described by the
introduction of cosmological constant \cite{Weinberg-RMP} that
determines the energy density $\rho_\Lambda=\mu_\Lambda^4$ at the
artificially small scale
\begin{equation}\label{untro1}
    \mu_\Lambda\approx 0.25\cdot 10^{-11}\;\mbox{GeV.}
\end{equation}
Indeed, due to the vacuum (zero-point) modes of quantum fields one
could expect the energy density
\begin{equation}\label{z5}
    \rho=\sum\limits_{\mbox{\tiny modes}}
    (-1)^F\hat \rho, \quad\mbox{with}\quad
    \hat \rho=\frac{1}{2}\int
    \frac{{\rm d}^3\boldsymbol k}{(2\pi)^3}\;\omega(\boldsymbol
    k),
\end{equation}
where $F=\{0,1\}$ denotes the fermion number for bosonic or
fermionic mode, correspondingly, while  the dispersion law is
given by $\omega(\boldsymbol k)=\sqrt{m^2+\boldsymbol k^2}$ in the
free-field approximation. So, exact supersymmetry (SUSY)
guarantees the balance of bosonic and fermionic modes with
$\omega>0$ as well as the equality of masses for superpartners,
i.e.
    $$
    I_W\hskip-2pt=\hskip-4pt\sum\limits_{\mbox{\tiny modes}}(-1)^F=0,\quad
    \{m_{\mbox{\footnotesize\textsc{b}}}=m_{\mbox{\footnotesize\textsc{f}}}
    \;\Leftrightarrow\;
    \omega_{\mbox{\footnotesize\textsc{b}}}(\boldsymbol k)=
    \omega_{\mbox{\footnotesize\textsc{f}}}(\boldsymbol k)\}.
    $$
Hence, the supersymmetric vacuum state
$|\Phi_{\mbox{\footnotesize\textsc{s}}}\rangle$ has zero energy
density $\rho_{\mbox{\footnotesize\textsc{s}}}=0$, and
$\mu_\Lambda\mapsto 0$ would be natural. The Witten's index $I_W$
\cite{Witten_index} would differ from zero in the supersymmetric
theory \cite{Weinberg-VIII}, if one introduces different numbers
of bosonic and fermionic modes with zero energy $\omega=0$,
but such the situation would correspond to
the case, when due to the conservation law for the number of
unpaired zero-energy modes, the supersymmetry cannot be
spontaneously broken in evident contradiction with observations.

A loss of balance in the dispersion laws of modes produces a
non-zero cosmological constant, when SUSY is broken down as we
observe in practice. Then, different masses are generated for
superpartners at scales below the characteristic energy of SUSY
breaking $\mu_{\mbox{\footnotesize\textsc{x}}}$, while SUSY is
restored at scales higher than
$\mu_{\mbox{\footnotesize\textsc{x}}}$. Therefore, the integration
in the energy density of single zero-point mode $\hat \rho$ in
(\ref{z5}) is actually cut off by
$\mu_{\mbox{\footnotesize\textsc{x}}}$ because of exact cancelling
by the superpartner contribution at higher scales\footnote{See
notes on the scheme of regularization in \cite{ACGKS}.}.

In the matter sector various scenarios of breaking down SUSY
\cite{Weinberg-VIII} give masses $m$ essentially less than
$\mu_{\mbox{\footnotesize\textsc{x}}}$ in the observable sector of
the Standard Model, while in a hidden sector, vacuum expectations
values (\textsc{vev}) of auxiliary scalar fields produce masses
$m_{\mbox{\footnotesize\textmd{hid.}}}\sim\mu_{\mbox{\footnotesize\textsc{x}}}$.
So, the contribution of observable-matter sector to vacuum energy
is suppressed with respect to
$\mu_{\mbox{\footnotesize\textsc{x}}}^4$ at least by powers of
ratio $m/\mu_{\mbox{\footnotesize\textsc{x}}}$. In contrast, the
hidden sector provides the vacuum energy about $\rho\sim\pm
\mu_{\mbox{\footnotesize\textsc{x}}}^4$.

In the gravity sector, SUSY breaking leads to the massive
goldstino with spirality $\pm\frac{1}{2}$ complementing the
gravitino modes with spirality $\pm\frac{3}{2}$ to the full set,
while the gravitino-superpartner, i.e., the graviton with
spirality $\pm2$ remains massless. Therefore, the goldstino breaks
the balance in the gravity sector, and the vacuum energy gains the
large negative contribution of two goldstino-modes with the mass
essentially less than the scale of SUSY breaking
\cite{Weinberg-VIII}
\begin{equation}\label{z-grav}
    \sum\limits_{\mbox{\tiny gravity}}(-1)^F\hat\rho\approx
    -\hskip-2mm\sum\limits_{\mbox{\tiny
    goldstino}}\hskip-2mm\hat\rho\approx
    -\frac{1}{8\pi^2}\,\mu_{\mbox{\footnotesize\textsc{x}}}^4,
\end{equation}
to the leading order in
${m}/{\mu_{\mbox{\footnotesize\textsc{x}}}}$. However, such the
disbalance in Wittens's index for gravity modes considers to be
artificial in global sense, since the glodstino is formed by a
superposition of spinor modes from the hidden sector, or
equivalently the spinor modes of superfields with non-zero
\textsc{vev} for their auxiliary scalar fields are expanded in
terms of goldstino and other unitary fields, whereas the
coefficients of expansion are given by the \textsc{vev}
\cite{Weinberg-VIII}. Then, the goldstino has two
bosonic-superpartner modes in the hidden sector, which produce a
positive contribution to the vacuum energy.
Nevertheless, Eq.(\ref{z-grav})
demonstrates a possibility to get the definite sign of vacuum
energy, which is \textit{negative}. More strictly, W.\hskip3pt
Nahm has algebraically found \cite{Nahm} that SUSY is forbidden in
four-dimensional (4D) spacetime with a positive density of vacuum
energy, while it is permitted in 4D spacetime with a negative
density of vacuum energy.

Thus, the vacuum modes in supergravity with SUSY broken below
$\mu_{\mbox{\footnotesize\textsc{x}}}$ give the \textit{negative}
cosmological constant, that corresponds to Anti-de Sitter
spacetime (AdS). We denote such the state by
$|\Phi_{\mbox{\footnotesize\textsc{x}}}\rangle$ with
$\rho=-\rho_{\mbox{\footnotesize\textsc{x}}}\sim-
\mu_{\mbox{\footnotesize\textsc{x}}}^4$ naturally setting
$\mu_{\mbox{\footnotesize\textsc{x}}}\gg \mu_\Lambda$ in
phenomenology of particle physics.

However, since the zero-point modes with momenta greater than
$\mu_{\mbox{\footnotesize\textsc{x}}}$ are common for both
$|\Phi_{\mbox{\footnotesize\textsc{s}}}\rangle$ and
$|\Phi_{\mbox{\footnotesize\textsc{x}}}\rangle$ states, they are
not completely independent. Hence, dynamical processes at
characteristic distances less than
$\lambda_{\mbox{\footnotesize\textsc{x}}}=1/\mu_{\mbox{\footnotesize\textsc{x}}}$
involve the correlation of two vacuum-states with zero and
negative cosmological constants.

The overlapping of two vacua is associated with a domain wall
separating the bubble of lower energy AdS-vacuum from the exterior
of higher energy flat-vacuum. A decay of unstable false vacuum
into the stable state \cite{false-decay} is described in terms of
\textit{bounce} being the solution of 4D-Euclidean
spherically-symmetric field-equations for a scalar field
interpolating between local minima of its potential in the region
of domain wall. The bounce determines the quasiclassical exponent
of penetration between two levels of vacuum. Coleman and De Luccia
\cite{CdL} shown that the bounce is essentially modified by
gravity that introduces a critical surface tension of domain wall,
while S.Weinberg \cite{Wein-S} found that the real surface density
of energy exceeds the critical one in supergravity. Thus, the
decay does not take place\footnote{See some further arguments in
\cite{Banks-Heretics}.}. Therefore, we focus on stationary 3D
spherically symmetric fluctuations of scalar field, that provide
the \textit{mixing of two vacuum-states}, if such the domain wall
cannot evolve to spatial infinity.

In this letter we find that the correlation of two states,
$|\Phi_{\mbox{\footnotesize\textsc{s}}}\rangle$ and
$|\Phi_{\mbox{\footnotesize\textsc{x}}}\rangle$ corresponding to
the exact and broken SUSY, leads to a fine mixing that produces
the stationary vacuum with a naturally small positive cosmological
constant related with the scale of SUSY breaking
$\mu_{\mbox{\footnotesize\textsc{x}}}$.


Consider action $S$ for fields independent of time. Then, we can
introduce the static potential $U^{\rm stat}$ by expression
\begin{equation}\label{stat1}
    S=\int\mathscr{L}\,\sqrt{-g}\;\;{\mathrm d}^4x\;
    \mapsto\;  S^{\rm stat}=   -U^{\rm stat}\int{\mathrm d}t,
\end{equation}
while the metric takes the form consistent with the spherical
symmetry
\begin{equation}\label{stat2}
    {\mathrm d}s^2=\widetilde{\mathtt{B}}(r)\,{\mathrm
    d}t^2-\frac{1}{\mathtt{B}(r)}\,{\mathrm d}r^2-r^2({\mathrm
    d}\vartheta^2-\sin^2\vartheta\,{\mathrm d}\varphi^2).
\end{equation}
The lagrangian of real scalar
$\mathscr{L}_f=\frac{1}{2}\,g^{\mu\nu}\partial_\mu\phi
\partial_\nu\phi-V(\phi)$ includes the gradient term
$
    g^{\mu\nu}\partial_\mu\phi
    \partial_\nu\phi\; \mapsto\;
    -(\phi^\prime)^2\,\mathtt{B},
$ where the prime denotes the derivative with respect to the
distance $r$. Then, the field equation reads as follows
\begin{equation}\label{stat4}
    \phi^{\prime\prime}+
    u^\prime\phi^\prime+\frac{2}{r}\,\phi^\prime=\frac{1}{\mathtt{B}}\;
    \frac{\partial V}{\partial\phi}
\end{equation}
with $u=\frac{1}{2}\ln(\widetilde{\mathtt{B}}\mathtt{B})$.

The energy-momentum tensor
$T_{\mu\nu}=\partial_\mu\phi\partial_\nu\phi-g_{\mu\nu}\mathscr{L}_f$
is composed by diagonal elements
\begin{equation}\label{stat5}
    \begin{array}{rcl}
    T_t^t&=& +
    \frac{1}{2}\,(\phi^\prime)^2\,\mathtt{B}+V,\\[2mm]
    T_r^r&=&-\frac{1}{2}\,(\phi^\prime)^2\,\mathtt{B}+V,
    \end{array}
\end{equation}
and $T_\vartheta^\vartheta=T_\varphi^\varphi=T_t^t$, which enter
the Einstein equations $ R_{\mu\nu}-\frac{1}{2}\,g_{\mu\nu}R=8\pi
G\,T_{\mu\nu}$.

Then, summing up the static actions of general relativity with the
lagrangian $\mathscr{L}_{GR}=-R/16\pi\,G$ and of static field with
the lagrangian $\mathscr{L}_f=-T_t^t$ under the relation of scalar
curvature with the trace of energy-momentum tensor $R=-8\pi G\,T$,
we get $U^{\rm stat}$ depending on the size of sphere $r_A$ inside
of which the matter has a non-zero energy,
\begin{equation}\label{stat6}
    U^{\rm stat}(r_A)=-4\pi\int\limits_0^{r_A}    V(\phi)\,
    \sqrt{\frac{\widetilde{\mathtt{B}}}{\mathtt{B}}}\;r^2\,{\mathrm d}r.
\end{equation}

The static potential $U^{\rm stat}$ equals zero if the scalar
field is global and it is positioned at a local minimum of its
potential with $V=0$. If the local minimum has negative
$V=-\rho_{\mbox{\footnotesize\textsc{x}}}$, then we arrive to
Anti-de Sitter spacetime with
\begin{equation}\label{AdS1}
    \widetilde{\mathtt{B}}_{\rm AdS}=\mathtt{B}_{\rm AdS}=1+\frac{r^2}{\ell^2},\qquad
    \frac{1}{\ell^2}=\frac{8\pi\,G}{3}\,\rho_{\mbox{\footnotesize\textsc{x}}},
\end{equation}
and the \textit{positive} static potential
\begin{equation}\label{AdS2}
    U^{\rm stat}_{{\rm AdS}}=\frac{4\pi}{3}\,r_A^3\,\rho_{\mbox{\footnotesize\textsc{x}}}.
\end{equation}

Let $\phi(r)$ be the solution, which interpolates between two
local minima of potential with zero energy and negative
$V=-\rho_{\mbox{\footnotesize\textsc{x}}}$. In this letter we
restrict ourselves by the consideration of thin domain wall, so
that the field is essentially changing in a narrow layer of width
$\delta r$ near the sphere of radius $r_A$ and $\delta r\ll r_A$.
Then,
\begin{equation}\label{stat7}
    U^{\rm stat}(r_A)=\frac{4\pi}{3}\,r_A^3\,\rho_{\mbox{\footnotesize\textsc{x}}}
    -4\pi\,r_A^2\,W_A,
\end{equation}
where $W_A$ determines the surface energy per unit area
\begin{equation}\label{stat7a}
    W_A(r_A)=\frac{1}{r_A^2}\int\limits_{r_A}^{r_A+\delta r}    V(\phi)\,
    \sqrt{\frac{\widetilde{\mathtt{B}}}{\mathtt{B}}}\;r^2\,{\mathrm d}r,
\end{equation}
and it is positive if the local minima are separated by
sufficiently high potential barrier.

For instance, at $\delta r\ll r_A\ll \ell$ we can safely neglect
the contribution of first-derivative terms in the field equation
(\ref{stat4}), since by the order of magnitude
$\phi^{\prime\prime}\sim \delta \phi/(\delta r)^2$, while the
spatial term is at the level of $\phi^\prime/r\sim\delta
\phi/(\delta r)^2\cdot \delta r/r_A\ll \phi^{\prime\prime}$, and
the metric elements $\widetilde{\mathtt{B}}$, $\mathtt{B}$ are
infinitely close to unit, so that $u^\prime\phi^\prime\sim
r_A^2/\ell^2\cdot 1/\delta r\cdot\delta\phi/\delta r\ll
\phi^{\prime\prime}$. Therefore, in this limit the field equation
does not involve any scale parameter external with respect to the
potential $V$ and it reproduces the ``kink'' solution with the
small value of $r_A$ and the width $\delta r$ determined by a mass
parameter in $V$, since the field equation yields $1/(\delta
r)^2\sim \delta V/(\delta\phi)^2\sim\partial^2 V/\partial \phi^2$.
Note, that the gradient contribution to the energy density $T_t^t$
equals the potential term. The kink sets the distribution of
matter determining the behavior of metric. Thus, the thin domain
wall can be established in the limit of small bubble, at least.
The domain wall can remain thin at $r_A\sim \ell$ or $r_A\gg \ell$
if the gravitational term in the field equation is suppressed, and
hence, $W_A\approx \mbox{const.}$

Setting $\widetilde{\mathtt{B}}\sim\mathtt{B}$ and $V\sim
(\phi^\prime)^2$, we roughly find
$W_A\sim\int\sqrt{V}\phi^\prime{\rm d}r\sim\int\sqrt{V}{\rm
d}\phi$, while in the supersymmetric theory with the chiral
superfield the potential is determined by the superpotential $f$
as $V=|\partial f/\partial \phi|^2$, hence, $W_A\sim |f_0|$, where
$f_0$ is the superpotential value at the vacuum. In supergravity
the \textit{negative} vacuum energy at the extremal of
superpotential is assigned to the superpotential itself in the
linear order in Newton constant $G$
\begin{equation}\label{sg1}
    \rho_{\mbox{\footnotesize\textsc{x}}}=24\pi G\,|f_0|^2,
\end{equation}
that yields
\begin{equation}\label{w1}
    W_A\sim m_{\mathtt{Pl}}\,\mu_{\mbox{\footnotesize\textsc{x}}}^2,
\end{equation}
where $m_{\mathtt{Pl}}=1/\sqrt{G}\sim 10^{19}$ GeV is the Planck
mass.

The materialization of bubble with zero static potential can take
place in the vacuum with zero density of energy at the
characteristic size given by solving $U^{\rm stat}=0$, hence,
\begin{equation}\label{size}
    r_A=\frac{3W_A}{\rho_{\mbox{\footnotesize\textsc{x}}}}\sim \ell.
\end{equation}
The materialization of bubble in the flat vacuum results in the
instability, since it takes place at the size $r_A$, that is not
positioned at the local minimum of static potential: \textit{the
domain wall begins to move to the bubble center} in agreement with
(\ref{stat7}). Furthermore, the zero size of bubble is also
unstable: the flat vacuum suffers from fluctuations due to the
bubbles of AdS vacuum.

This situation is opposite to the case of \textit{switching off
the gravity}, when the domain wall can materialize after the
penetration through the potential barrier, so that it will move to
spatial infinity, that means the \textit{decay} of flat vacuum to
the AdS one.

As we have just shown the gravity induces the materialization of
3D spherical bubble not propagating to infinity, that means the
\textit{mixing} of two levels, but \textit{not the decay}. Thus,
due to the unstable bubbles the vacua are not eigenstates of true
hamiltonian.

Let us consider the quantum system of two stationary vacuum-levels
described by the hamiltonian density in the volume restricted by
the domain wall,
\begin{equation}\label{2l-1}
\begin{array}{rl}
    \mathscr{H}=&-\rho_{\mbox{\footnotesize\textsc{x}}}|\Phi_{\mbox{\footnotesize\textsc{x}}}\rangle\langle\Phi_{\mbox{\footnotesize\textsc{x}}}|+
    \rho_{\mbox{\footnotesize\textsc{s}}}|\Phi_{\mbox{\footnotesize\textsc{s}}}\rangle
    \langle\Phi_{\mbox{\footnotesize\textsc{s}}}|\\[3mm]
    &
    +\widetilde\rho\,\big\{
    |\Phi_{\mbox{\footnotesize\textsc{x}}}\rangle
    \langle\Phi_{\mbox{\footnotesize\textsc{s}}}|+
    |\Phi_{\mbox{\footnotesize\textsc{s}}}\rangle
    \langle\Phi_{\mbox{\footnotesize\textsc{x}}}|
    \big\},
\end{array}
\end{equation}
where
$\rho_{\mbox{\footnotesize\textsc{x}}}\sim\mu_{\mbox{\footnotesize\textsc{x}}}^4$
in the AdS vacuum with broken SUSY, while in the supersymmetric
vacuum $\rho_{\mbox{\footnotesize\textsc{s}}}=0$, and
$\widetilde\rho$ takes a real positive value due to the freedom in
the definition of vacuum states. Let us, first, evaluate the width
of domain wall $\delta r$  and, second, estimate the mixing matrix
element $\widetilde\rho=
\langle\Phi_{\mbox{\footnotesize\textsc{s}}}|\mathscr{H}|
\Phi_{\mbox{\footnotesize\textsc{x}}}\rangle$.

If the domain wall is thin, its mass is given by the expression
$M_{\rm dw}=4\pi r_A^2\,W_A\sim 4\pi \ell^2 \delta r\,V_0$, where
$V_0$ is the characteristic height of potential barrier inside the
wall. This mass is compensated by the negative mass of bubble
$M_{\rm b}=-4\pi r_A^3 \rho_{\mbox{\footnotesize\textsc{x}}}/3\sim
-\mu_{\mbox{\footnotesize\textsc{x}}}^4\ell^3$, so that under
$\ell\sim m_{\mathtt{Pl}}/\mu_{\mbox{\footnotesize\textsc{x}}}^2$
we get
\begin{equation}\label{2l-2}
    \delta r\cdot V_0\sim \ell\,\rho_{\mbox{\footnotesize\textsc{x}}}\sim m_{\mathtt{Pl}}\,\mu_{\mbox{\footnotesize\textsc{x}}}^2.
\end{equation}
Furthermore, for the chiral superfield due to (\ref{sg1}) one
finds
\begin{equation}\label{2l-3}
    f_0\sim m_{\mathtt{Pl}}\,\mu_{\mbox{\footnotesize\textsc{x}}}^2\quad
    \Rightarrow\quad V_0\sim\frac{f_0^2}{(\delta\phi)^2}\sim
    \frac{m_{\mathtt{Pl}}^2\,\mu_{\mbox{\footnotesize\textsc{x}}}^4}{(\delta\phi)^2},
\end{equation}
where $\delta \phi$ is the characteristic change of field in the
domain wall, i.e. the ``distance'' between two extremal points of
the field. Hence, we evaluate the width of thin domain wall in
terms of evolution change of the field,
\begin{equation}\label{2l-4}
    \delta r\sim\frac{(\delta\phi)^2}{m_{\mathtt{Pl}}\,\mu_{\mbox{\footnotesize\textsc{x}}}^2}.
\end{equation}
Putting $\delta r\ll r_A\sim \ell$, we find
$
    \delta\phi\ll m_{\mathtt{Pl}}.
$ 
Therefore, the domain wall is thin, if the field dynamics is
essentially sub-Planckian. For instance, we get
\begin{eqnarray}
  \delta \phi\sim \mu_{\mbox{\footnotesize\textsc{x}}} &\Rightarrow& \delta r\sim\frac{1}{m_{\mathtt{Pl}}},
  \\[2mm]
  \delta\phi\sim\sqrt{m_{\mathtt{Pl}}\,\mu_{\mbox{\footnotesize\textsc{x}}}} &
  \Rightarrow&\delta
  r\sim\lambda_{\mbox{\footnotesize\textsc{x}}}=
  \frac{1}{\mu_{\mbox{\footnotesize\textsc{x}}}}.
\end{eqnarray}
The case of $\delta r\sim
\lambda_{\mbox{\footnotesize\textsc{x}}}$ looks the most natural
situation, since the domain wall has the size of correlation
length of two vacua. Note, that at $\sqrt{m_{\mathtt{Pl}}
\,\mu_{\mbox{\footnotesize\textsc{x}}}}\ll\delta \phi\ll
m_{\mathtt{Pl}}$ the width of thin domain wall becomes much
greater that the correlation length
$\lambda_{\mbox{\footnotesize\textsc{x}}}$, that requires especial
consideration, which can be presented elsewhere.

The correlation energy of two states can be estimated in terms of
mixing density of energy multiplied by the volume of the bubble,
\begin{equation}\label{2l-5}
    E_{\rm corr.}\sim \widetilde \rho\cdot\ell^3.
\end{equation}
On the other hand, it is determined by the energy in the
overlapping region restricted by the correlation length
$\lambda_{\mbox{\footnotesize\textsc{x}}}$, i.e. in the element of
thin domain wall with the area of the order of
$\lambda_{\mbox{\footnotesize\textsc{x}}}^2$. Hence, $E_{\rm
corr.}$ is given by the surface tension $W_A\sim \delta r\cdot
V_0$ in the area of correlation
\begin{equation}\label{2l-5a}
    E_{\rm corr.}\sim W_A\cdot \lambda_{\mbox{\footnotesize\textsc{x}}}^2.
\end{equation}
In other words, the correlation energy is determined by the mass
of domain wall with size
$\lambda_{\mbox{\footnotesize\textsc{x}}}$, that could be
considered as a bare AdS-bubble in the beginning of
materialization. Therefore, under $W_A\sim f_0\sim
m_{\mathtt{Pl}}\,\mu_{\mbox{\footnotesize\textsc{x}}}^2$ we get
the estimate
\begin{equation}\label{2l-6}
    \widetilde\rho\sim \frac{\mu_{\mbox{\footnotesize\textsc{x}}}^2}{\ell^2}\sim
    \frac{\mu_{\mbox{\footnotesize\textsc{x}}}^6}{m_{\mathtt{Pl}}^2},
\end{equation}
implying $\widetilde\rho\ll\rho_{\mbox{\footnotesize\textsc{x}}}$.
Then, the matrix of two-level hamiltonian of vacuum has the form
\begin{equation}\label{s-s1}
    \mathscr{H}=\left(\hskip-3pt%
\begin{array}{cc}
  -\rho_{\mbox{\footnotesize\textsc{x}}} & \widetilde\rho \\[2mm]
  \widetilde\rho & 0 \\
\end{array}%
\right) \quad\mbox{at}\quad
\widetilde\rho\ll\rho_{\mbox{\footnotesize\textsc{x}}},
\end{equation}
so that such the texture is well known in the particle
phenomenology as the ``seesaw mechanism'' for describing the
mixing of charged currents, for instance \cite{Fritzsch}.
{Some applications of seesaw mechanism to the cosmological
constant problem and quintessence dynamics have been
studied in \cite{Grav-seesaw} and
\cite{Enqvist:2007tb}.} 
The eigenvalues are equal to
\begin{equation}\label{s-s2}
    \rho_{\Lambda}=-\frac{1}{2}\left(\rho_{\mbox{\footnotesize\textsc{x}}}\pm\sqrt{\rho_{\mbox{\footnotesize\textsc{x}}}^2+4\widetilde\rho^2}\right),
\end{equation}
and due to
$\widetilde\rho\ll\rho_{\mbox{\footnotesize\textsc{x}}}$ they are
reduced to
\begin{equation}\label{s-s3}
    \begin{array}{lcc}
      \rho_\Lambda^{\rm dS} & \approx &\displaystyle\frac{\widetilde\rho^2}{\rho_{\mbox{\footnotesize\textsc{x}}}},
      \\[4mm]
      \rho_\Lambda^{\rm AdS} & \approx & -\rho_{\mbox{\footnotesize\textsc{x}}}, \\
    \end{array}
\end{equation}
that corresponds to expanding de Sitter universe and collapsing
AdS universe. Both vacua are stationary levels with no mixing or
decay. We are certainly living in the Universe with the dS vacuum.

The eigenstates are described by superposition of initial
non-stationary vacua
\begin{equation}\label{s-s4}
    \begin{array}{llcl}
      |\mbox{vac}\rangle & \hskip-3pt= \cos\theta_{\mbox{\footnotesize\textsc{k}}} |\Phi_{\mbox{\footnotesize\textsc{s}}}\rangle
      &\hskip-5pt+&\hskip-3pt\sin\theta_{\mbox{\footnotesize\textsc{k}}} |\Phi_{\mbox{\footnotesize\textsc{x}}}\rangle,\\[2mm]
      |\mbox{vac}'\rangle & \hskip-3pt= \cos\theta_{\mbox{\footnotesize\textsc{k}}} |\Phi_{\mbox{\footnotesize\textsc{x}}}\rangle
      &\hskip-5pt-&\hskip-3pt\sin\theta_{\mbox{\footnotesize\textsc{k}}} |\Phi_{\mbox{\footnotesize\textsc{s}}}\rangle,
    \end{array}
\end{equation}
with the mixing angle equal to
\begin{equation}\label{s-s5}
    \tan 2\theta_{\mbox{\footnotesize\textsc{k}}}
    =\frac{2\widetilde\rho}{\rho_{\mbox{\footnotesize\textsc{x}}}},
\end{equation}
well approximated by
\begin{equation}\label{s-s5a}
    \sin\theta_{\mbox{\footnotesize\textsc{k}}}\approx
    \frac{\widetilde\rho}{\rho_{\mbox{\footnotesize\textsc{x}}}}\ll 1.
\end{equation}
The thin domain wall determines
\begin{equation}\label{e1}
    \rho_\Lambda^{\rm dS}\sim
    \frac{\mu_{\mbox{\footnotesize\textsc{x}}}^8}{m_{\mathtt{Pl}}^4},
\end{equation}
and due to $\rho_\Lambda=\mu_\Lambda^4$ we get the estimate
\begin{equation}\label{e2}
    \mu_{\mbox{\footnotesize\textsc{x}}}\sim\sqrt{m_{\mathtt{Pl}}\,\mu_\Lambda}\sim 10^4\,\mbox{GeV}.
\end{equation}
Thus, the thin domain wall is relevant to the low scale of SUSY
breaking. Note, that estimate (\ref{e2}) was obtained by T.Banks
in \cite{Banks-I} in other way of physical argumentation for the
mechanism of SUSY breaking.

The relation of SUSY breaking scenario with different regimes of
domain wall fluctuations can be clarified by considering some
typical properties of scalar field potential, that would be
presented elsewhere \cite{in_preparation}.

For thin domain wall, the mixing angle of two levels
$\theta_{\mbox{\footnotesize\textsc{k}}}$ takes the value about
\begin{equation}\label{t1}
    \theta_{\mbox{\footnotesize\textsc{k}}}\approx
    \frac{\widetilde\rho}{\rho_{\mbox{\footnotesize\textsc{x}}}}\sim
    \frac{\mu_{\mbox{\footnotesize\textsc{x}}}^2}{m_{\mathtt{Pl}}^2}\sim\frac{\mu_\Lambda}{m_{\mathtt{Pl}}}.
\end{equation}
Therefore, it is certainly fixed by present day data on the
cosmological constant,
$\theta_{\mbox{\footnotesize\textsc{k}}}\sim 10^{-30}$.

In conclusion, we have described the mechanism for dynamical
generation of small cosmological constant due to seesaw mixing of
two initial vacuum-states describing the phases of exact and
broken supersymmetry. The current value of cosmological constant
is consistent with phenomenological estimates of SUSY broken scale
in particle physics. The mechanism works due to fluctuations
formed by bubbles of AdS vacuum separated by thin domain walls
from the flat vacuum.

\vspace{2mm} The work of V.V.K. is partially supported by the
Russian Foundation for Basic Research, grant 07-02-00417.

\end{document}